# Transparent Multispectral Photonic Electrode for All-Weather Stable and Efficient Perovskite Solar Cells

George Perrakis,[1,*] Anna C. Tasolamprou,[2] George Kakavelakis,[3,4,*] Konstantinos Petridis,[3] Michael Graetzel,[4] George Kenanakis,[1] Stelios Tzortzakis,[1,5,6] Maria Kafesaki[1,5]

[1]Institute of Electronic Structure and Laser (IESL), Foundation for Research and Technology - Hellas (FORTH), 70013 Heraklion, Crete, Greece

[2]Department of Physics, Section of Electronic Physics and Systems, National and Kapodistrian University of Athens, 15784, Athens, Greece

[3]Department of Electronic Engineering, Hellenic Mediterranean University, Romanou 3, Chalepa, GR-73100, Chania, Crete, Greece

[4]Laboratory of Photonics and Interfaces, Institute of Chemical Sciences and Engineering, Ecole Polytechnique Fédérale de Lausanne, 1015, Lausanne, Switzerland

[5]Department of Materials Science and Technology, University of Crete, 70013 Heraklion, Crete, Greece

[6]Texas A&M University at Qatar, 23874 Doha, Qatar

*Corresponding Authors: gperrakis@iesl.forth.gr, kakavelakis@hmu.gr

**ABSTRACT:** Perovskite solar cells (PSCs) are the most promising technology for advancing current photovoltaic performance. However, the main challenge for their practical deployment and commercialization is their operational stability, affected by solar illumination and heating, as well as the electric field that is generated in the PV device by light exposure. Here, we propose a transparent multispectral photonic electrode placed on top of the glass substrate of solar cells, which simultaneously reduces the device solar heating and enhances its efficiency. Specifically, the proposed photonic electrode, composed of a low-resistivity metal and a conductive layer, simultaneously serves as a highly-efficient infrared filter and an ultra-thin transparent front contact, decreasing devices' solar heating and operating temperature. At the same time, it simultaneously serves as an anti-reflection coating, enhancing the efficiency. We additionally enhance the device cooling by coating the front glass substrate side with a visibly transparent film (PDMS), which maximizes substrate's thermal radiation. To determine the potential of our photonic approach and fully explore the cooling potential of PSCs, we first provide experimental characterizations of the absorption properties (in both visible and infrared wavelengths) of state-of-the-art PSCs among the most promising ones regarding the efficiency, stability, and cost. We then numerically show that applying our approach to promising PSCs can result in lower operating temperatures by over 9.0 °C and an absolute efficiency increase higher than 1.3%. These results are insensitive to varying environmental conditions. Our approach is simple and only requires modification of the substrate (fabrication substrate becomes top-layer at the operation state) in the conventional next-generation solar cell fabrication process; it therefore points to a feasible photonic approach for advancing current photovoltaic performance with next-generation solar cell materials. Additionally, besides solar cells, our approach can be applied to heat-insulating or other energy-saving systems, such as smart windows or hybrid thermoelectric-photovoltaic systems.

■ **INTRODUCTION**

Perovskite solar cells (PSCs) are the most promising technology to hit the market among the third generation of photovoltaics (PVs).[1] Their optimal band gap (∼1.4–1.5 eV), coupled with exceptional photovoltaic performance (26 %) and low manufacturing cost, is attractive for advancing the current PV conversion of solar energy. However, their limited stability hinders commercial exploitation. Stability studies on PSCs have highlighted the detrimental effect of external stimuli such as moisture, oxygen, and ultraviolet light on the device's stability without appropriate encapsulation.[2] However, encapsulated PSCs still degrade under illumination, leading to unsatisfactory operational stability, especially compared to commercial inorganic-based solar cells.[3,4] Specifically, it was shown that elevated device temperature coupled with excess charge carriers and the electric field generated under constant illumination is the dominant force in their rapid degradation.[5] Power output degradation in PSCs (e.g., due to electric field and diffusion driven ion migration, phase segregation, metastable crystal transition, volatilization, metal diffusion) significantly accelerates at elevated temperatures (>40 °C).[2-6] Under natural conditions of outdoor operation, the device can easily reach temperatures higher than 50 °C, sometimes even reaching values close to the lamination-fabrication temperatures, which are carefully selected to be the lowest possible (<85 °C).[1] This heating of the perovskite solar cell has adverse consequences on its performance and reliability. Besides reduced lifetime, the power conversion efficiency (PCE) also decreases with increasing temperature. For PSCs, every 1 °C temperature rise leads to a relative efficiency decline of about 0.21%.[1]

Heat is produced within the module due to photo-generated carriers thermalization, recombination, and parasitic absorption of incident photons in the various parts/layers of the solar cell.[7] A number of strategies based on passive methods for solar module cooling have been proposed to mitigate this problem, including infrared filter designs that selectively reflect sub-bandgap radiation that does not contribute to the generation of photocurrent (∼$\lambda_g$–4 μm, where $\lambda_g$ is semiconductor's band gap wavelength)[8,9] or optical designs that increase the device emissivity in the mid-infrared range within the atmospheric transparency window and therefore enhance radiative cooling of the module.[7,10-14]

In our study, we focus on a more direct way which addresses the root cause of temperature rise by identifying and mitigating heat sources. Specifically, optical simulation studies in next-generation solar cells, such as perovskite- or organic-based thin-film solar cells, show that parasitic absorption mainly occurs in the conventional transparent conductive oxide (TCO)-type front-contact electrodes [such as fluorine-doped tin oxide (FTO, $SnO_2$:F) or tin-doped indium oxide (ITO, $SnO_2$: $In_2O_3$)].[15] This absorption does not contribute to the generation of photocurrent and acts as a heat source, seriously affecting performance and reliability. The reason is the well-known trade-off relationship in front-contact electrodes between the sheet resistance ($R_{sh}$ – which contributes significantly to the distributed series resistance or the equivalent resistance at the maximum power point of a solar cell, $R_s$) and the transmittance ($T$ – which contributes to the photocurrent, $J_{ph}$). As a result, a front-contact electrode is expected to significantly affect the solar cell's operating temperature as a result of solar radiation heating,[16] Joule heating,[17] and non-radiative carrier recombination.[18]

Numerous novel front-contact electrodes emerged in recent years to mitigate the $T$-$R_{sh}$ trade-off effect,[16,19,20] enhance the system's operational stability,[21] or reduce the cost.[17,21] Promising candidates are (i) TCO electrodes with higher transparency at a given conductivity, such as indium-doped tin oxide (ITO) or aluminum-doped zinc oxide (AZO, ZnO:Al), (ii) metal-based electrodes, such as Ag nanowire transparent thin-films (AgNW),[17] U-shaped gold layer on top of the TCO,[16] and TCO/metal/TCO composite structures,[19,20] or (iii) carbon-based electrodes, such as carbon nanotubes (CNTs) and reduced graphene oxide (RGO).[21] Despite being demonstrated as a promising way to mitigate the $T$-$R_{sh}$ trade-off, parasitic absorption of incident photons in the other parts/layers of the solar cell apart from the front contact[15] also leads to increased solar heating in the cell.

In this paper, we propose a transparent multispectral photonic electrode placed on top of the glass substrate of solar cells for optimal thermal management and PCE enhancement. Specifically, in contrast to conventional electrodes, which are only responsible for the collection of photo-generated carriers, the proposed photonic electrode simultaneously serves as a highly-efficient infrared filter and an ultra-thin transparent front contact, decreasing devices' solar heating and operating temperature. At the same time, it simultaneously serves as an anti-reflection coating, enhancing the PCE. Specifically, the electrode consists of a low-resistivity metal and a conductive layer, with an overall thickness lower than 70 nm, which are encapsulated by one or more common-oxide or dielectric layers. In contrast to common infrared filter designs,[9] the proposed structure requires a low number of layers for similar reflection amplitude and bandwidth, owing to the integration of the reflective metal layer. Moreover, it does not affect substrate's thermal emission in the atmospheric transparency window in mid-infrared,[10-14] maintaining high-efficiency cooling and compactness. To additionally consider this effect and further enhance the device cooling, we coat solar cells' glass substrate (front side) with a visibly transparent film, which maximizes substrate's thermal radiation. To determine the potential of our approach and fully exploit the cooling potential of PSCs, we first provide experimental characterizations of the absorption properties of the most promising (regarding the efficiency, stability, and cost) PSCs in the entire electromagnetic spectrum, in visible and infrared wavelengths. We then numerically show that applying our approach to PSCs can result in lower operating temperatures by over 9.0 °C and an absolute efficiency increase higher than 1.3%. These results are insensitive to varying environmental conditions due to enhanced selective-spectral, radiative cooling, and solar absorption in perovskite for all-weather stability and high performance. Our approach is simple and only requires modification of the substrate (fabrication substrate becomes top-layer at the operation state) in the conventional next-generation solar cell fabrication process and therefore points to a feasible photonic approach for advancing current photovoltaic performance with next-generation solar cell materials.

## ■ EXPERIMENTAL CHARACTERIZATION OF THE ABSORPTIVITY/EMISSIVITY SPECTRA OF PROMISING PEROVSKITE SOLAR CELLS IN THE VISIBLE AND INFRARED WAVELENGTH RANGE

To determine the potential of our photonic approach and fully explore the cooling potential of PSCs, we first experimentally examine their solar absorption and thermal emission properties. We consider two types of state-of-the-art PSCs with typical mesoporous architecture geometry: cell denoted by "w/ Au" is a PSC with Au back contacts for enhanced efficiency (Figure 1b), and cell denoted by "w/ MLG" is a PSC with a multilayer Graphene back contact for enhanced stability and low cost (Figure 1a). Specifically, cell configuration w/ Au exhibits record efficiencies, even beyond 25% for smaller cells,[22,23] but in expense to the increased cost, unsatisfactory operational stability, and limited large-scale application, hindering commercial exploitation.[24,25] Cell configurations w/ MLG exhibit lower efficiencies (~20%),[25,26] mainly due to open-circuit voltage ($V_{OC}$) loss as a result of hole trapping due to poor contact (pinholes and gaps) at the hole-transporting layer (HTL)-MLG interface (see the left panel in Figure 1d),[26] but much higher device operational stability and much lower cost due to high-throughput fabrication, e.g., by utilizing printing techniques and low-cost materials.[24,25,27] Moreover, we use highly-efficient low-band gap formamidinium lead iodide perovskite ($FAPbI_3$) with enhanced thermal stability as the photoactive layer for both solar cells. Specifically, until recently, methylammonium lead trihalide ($CH_3NH_3PbI_3$ or $MAPbI_3$) perovskite was extensively examined for PSCs due to its fabrication simplicity and its direct band-gap of ~1.55 eV (~800 nm, i.e., at the onset of the optical range), very close to the ideal compared to other perovskites in which another halide is present. Recently, $FAPbI_3$ has endowed PCEs up to 26% because of its optimal bandgap, enhanced transport properties, and thermal stability.[22,23] Indicatively, studies have utilized $FAPbI_3$-based PSCs with active layer thicknesses as high as 800 nm (compared to ~350 nm for $MAPbI_3$) without decreasing the external quantum efficiency, greatly enhancing PSCs' $J_{SC}$ and $V_{OC}$, hence PCE.[22,23] The rest of the structure is as follows: TEC 8 (FTO – 500 nm)-covered glass substrate (2.2 mm)/compact $TiO_2$ ($cTiO_2$ – 30 nm)/mesoporous $TiO_2$ ($mTiO_2$ – 200 nm)/perovskite ($FAPbI_3$ – 800 nm)/Spiro-OMeTAD (200 nm)/Au (80 nm) or multi-layer Graphene (10 μm). The transparent conducting oxide (FTO: fluorine doped tin oxide), multilayer Graphene (MLG), and gold (Au) layers are the front and rear electrical contacts, respectively, while the transparent soda-lime glass is the encapsulant substrate. Spiro-OMeTAD polymer is the hole-transporting layer (HTL) for both solar cells, while the compact ($cTiO_2$) and mesoporous ($mTiO_2$) titanium dioxide layers compose the electron-transporting layer (ETL). We characterized the structures with an FT-IR spectrometer and compared with the simulated results by employing the transfer matrix method (see the Methods section). To simulate the performance of our structures, we obtained the material parameters used for the active layer from ref 28, of the electrodes from refs 10,29–31, and of the other layers from refs 10,29,32. Examined PSCs' experimental and simulated solar absorption and thermal emissivity spectra are shown in Figure 1.

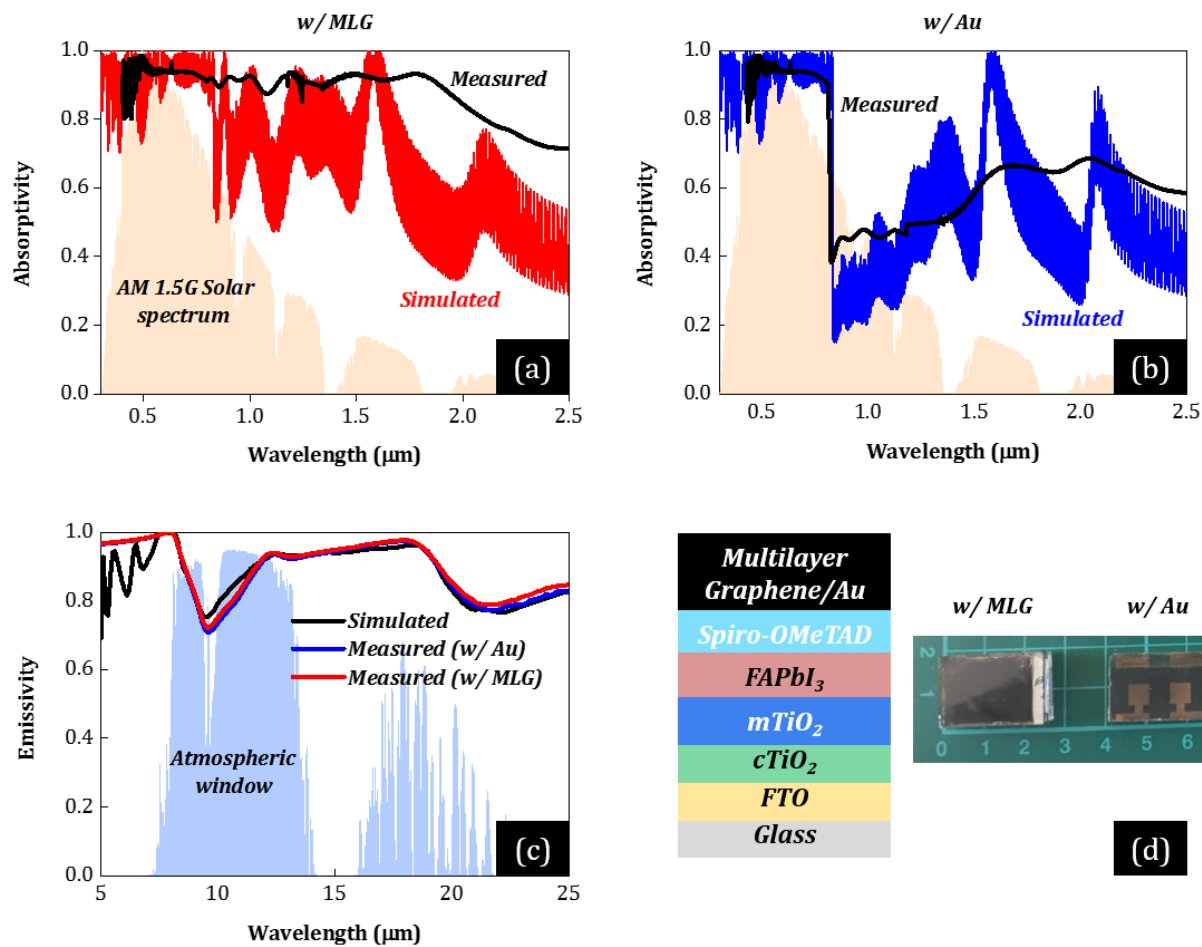

**Figure 1.** (a, b) Experimentally measured (black) and simulated (red and blue) solar absorption spectra of the two representative PSCs, cell w/ MLG (red) and w/ Au (blue), with the normalized AM1.5G solar spectrum plotted for reference (orange shaded area). Note that the PSC w/ Au absorbs much less light of one-micron wavelength than the PSC w/ MLG due to reflection from the device. (c) Experimentally measured (black) and simulated (red and blue) thermal emissivity spectra of the two representative PSCs, cell w/ MLG (red) and w/ Au (blue), with realistic atmosphere transmittance plotted for reference (blue shaded area). (d) Geometry of the PSCs investigated here (left). The role of the different layers of the solar cells is discussed in the main text. Photo of the back side of the two measured solar cells (right). Cell "w/ MLG": a perovskite solar cell with a multilayer Graphene back contact. Cell "w/ Au": a perovskite solar cell with Au back contacts.

In both PSC cases, good agreement between theory and experiment is observed, validating the numerical simulations. Deviations in the absorption spectra, i.e., less pronounced absorptivity peaks (originating from interference in the thin film stack) and a systematic higher absorptivity of the characterized samples compared to simulations can be attributed to the samples' roughness, whereas in the simulations planar interfaces were assumed. The absorptivity decrease at $\lambda > 1.9$ μm in all cases (measured and simulated) is attributed to the epsilon-near-zero (ENZ) response of conductive FTO in that regime.[29]

In the solar wavelength region (~0.3–4 μm, see Figure 1a and b), both PSCs show strong light absorption, especially at ~0.3–$\lambda_g$ μm, as expected. In the sub-band gap wavelength range, ~$\lambda_g$–4 μm, both PSCs show high absorption, despite sub-band gap photons' lower energy than active material's band gap energy ($E_g$). For the cell w/ Au, such strong sub-band gap absorption mainly originates from the FTO front contact,[15] representing a heat source. The sub-band gap absorption further increases due to interference in the thin-film stack. According to our calculations (i.e., from the simulated absorption in other layers than FAPbI$_3$ in 0.3–4 μm and the AM1.5G sun spectrum), the parasitic absorption in the cell w/ Au equals 200 W/m². This absorption does not contribute to photocurrent and acts as a heat source, seriously affecting solar cells' PCE and reliability. Compared to the cell w/ Au, the promising cell w/ MLG of low cost and higher device stability shows an even stronger absorption in the sub-band gap wavelength range (see red versus blue curve in Figure 1a), expected to lead to excess heat generation and even higher operating temperatures in realistic outdoor conditions. The reason is that multilayer Graphene strongly absorbs in the near-infrared regime compared to Au, which acts as a reflector. As a result, in the cell w/ MLG, parasitic absorption equals an even higher value of 310.6 W/m$^2$ (i.e., 110.6 W/m$^2$ more than the cell w/ Au). We note that PSCs' heat output is expected to be slightly higher than 200 and 310.6 W/m$^2$ due to higher experimental absorptivity than the simulations (see Figure 1a and b).

In mid-infrared wavelengths at ~4–25 μm (see Figure 1c), solar cells' front surface thermal emissivity comes almost exclusively from the millimeter-thick glass substrate. Additionally, PSC devices' emissivity is almost identical to all solar cell technologies (see Figure 1c) since glass is the most common/reliable encapsulant/substrate. These bulk layers have relatively strong thermal emissivity due to their phonon-polariton resonance modes at mid-infrared (MIR) wavelengths,[12,13] and hence have a cooling effect. However, the thermal emissivity of these layers is typically not optimized, especially within the atmospheric transparency window, which lowers their cooling power (see Figure 1c).

In summary, the results of Figure 1 indicate that most promising PSCs inevitably (to be efficient, cost-effective, and functional) have (i) strong parasitic absorption in the solar wavelengths and (ii) sub-optimal emissivity in the thermal wavelengths (~4–25 μm). As a result, even thin-film PSCs technology (~1 μm) suffers intense thermal loads on a sunny day,[1] comparable to bulkier silicon-based counterparts (~250 μm thickness),[33] seriously affecting their PCE and reliability.[3,5]

One way to reduce parasitic absorption in the cell w/ Au is to mitigate the $T$-$R_{sh}$ trade-off effect by utilizing an appropriate front TCO layer with higher transparency at a given conductivity compared to FTO. Common TCOs, though, such as optically superior indium-doped tin oxide (ITO) or aluminum-doped zinc oxide (AZO, ZnO:Al), limit PSC performance and application due to conductivity loss at high annealing temperatures required for the compact (cTiO$_2$) or mesoporous TiO$_2$ (mTiO$_2$) layers and increase the cost. Substituting cTiO$_2$ and mTiO$_2$ with other alternatives, such as PTAA or SnO$_2$, seems promising for utilizing various TCOs such as ITO,[34] reducing PSCs' solar heating due to lower parasitic absorption in the front contact. However, like the examined cell w/ Au (Figure 1b), such cells exhibit unsatisfactory operational stability, limited large-scale application, and high cost, mainly due to the metal back contact,[24] impeding commercial exploitation. Substituting the metal back contact of such cells with MLG to enhance their operational stability and reduce the cost leads to excess heat since sub-band gap radiation is mainly affected by the MLG, which strongly absorbs in the near-infrared regime. Moreover, the sheet resistance of conventional electrodes and TCOs (~8–15 Ω/sq) is still high when upscaling from a single cell to a module.[19] This results to lower module solar-to-electrical PCEs (due to less efficient carrier transport) and further increases heat dissipation, thus indicating that structural changes or other TCO candidates are required. These results indicate that a photonic electrode as discussed in the previous section, should highly impact PSCs PCE and reliability.

## ■ OUR APPROACH

We now summarize photonic electrodes' design requirements for lowering the device thermal load (maintaining high efficiency) and discuss the considerations to satisfy these requirements. Specifically, the photonic electrode should have high spectral selectivity and filtering properties in the entire solar wavelength range (0.3–4 μm). In the wavelength ranges of 0.3–0.385 μm and $\lambda_g$–4 μm (where $\lambda_g$ is perovskite's band gap wavelength, $\lambda_g$~0.832 μm for FAPbI$_3$), the photonic structure should have maximized reflection to reduce the parasitic heat generation and reflect the harmful ultraviolet (UV) radiation. In the wavelength range of 0.385–$\lambda_g$ μm where photons convert to photocurrent, the photonic electrode should have a minimum reflection. Moreover, the photonic electrode should have minimum parasitic absorption in the entire 0.3–4 μm solar wavelength range. Regarding solar cells' electrical properties, the photonic multilayer should also serve as a transparent front contact of low sheet resistance for efficient carrier collection.

We use a photonic approach for the multilayer/electrode to meet these demands. We propose to add a photonic multilayer consisting of two layers of silver (Ag) and molybdenum oxide (MoO$_x$) on top of four to eight alternating layers of titanium dioxide (TiO$_2$) and silicon dioxide (SiO$_2$) and two layers of aluminum oxide (Al$_2$O$_3$) and hafnium dioxide (HfO$_2$) of varying thicknesses, which are all deposited on the glass substrate (see Figure 2a). The top two layers (Ag and MoO$_x$) are primarily responsible for (i) efficient carrier transport and collection and (ii) broad-band reflection (from Ag) of the sub-band gap radiation. Specifically, encapsulated metal-based electrodes may achieve low sheet resistance (<8 Ω/sq),[19,20,35–39] where the resistivity may be tuned by varying the metal thickness. Moreover, due to silver's low resistivity and thickness (>8 nm) compared to conventional TCOs (~100 and 500 nm for ITO and FTO, respectively), they may effectively mitigate the $T$-$R_{sh}$ trade-off in solar cells due to simultaneous lower $R_{sh}$ and parasitic absorption. The bottom layers (alternating $TiO_2/SiO_2/HfO_2/Al_2O_3$) assist in (i) suppressing reflection from the metal (due to destructive interference of the reflected waves) and obtaining high transparency in the beneficial wavelength range, where photons convert to photocurrent (0.385–$\lambda_g$ μm), and (ii) optimizing solar reflection in the harmful UV (0.3–0.385 μm) and sub-band gap regime (~$\lambda_g$–4 μm), in a manner similar to that achievable using periodic 1D photonic crystals. $TiO_2$ and $HfO_2$ are the high-index materials,[10] while $SiO_2$ and $Al_2O_3$ are optically transparent and are the low-index and anti-reflection layers, respectively. To additionally enhance the device cooling without affecting solar cells' encapsulation properties and protection, we coat the glass substrate (front side) with a common inexpensive polydimethylsiloxane (PDMS) film, which maximizes glass's thermal radiation.

Consequently, the proposed photonic substrate affects the electrical and optical properties of the solar cell and its operating temperature and hence the cell's lifetime through a combination of material properties and interference effects, offering multifunctionality (see Figure 2a). Our approach is simple and only requires modification of the substrate (fabrication substrate becomes top-layer at the operation state) in the conventional next-generation solar cell fabrication process. Additionally, our approach is generic to solar cells with optimal band gap (~1.4–1.5 eV) or other heat-insulating and energy-saving devices such as smart windows and hybrid thermoelectric-photovoltaic systems.[40]

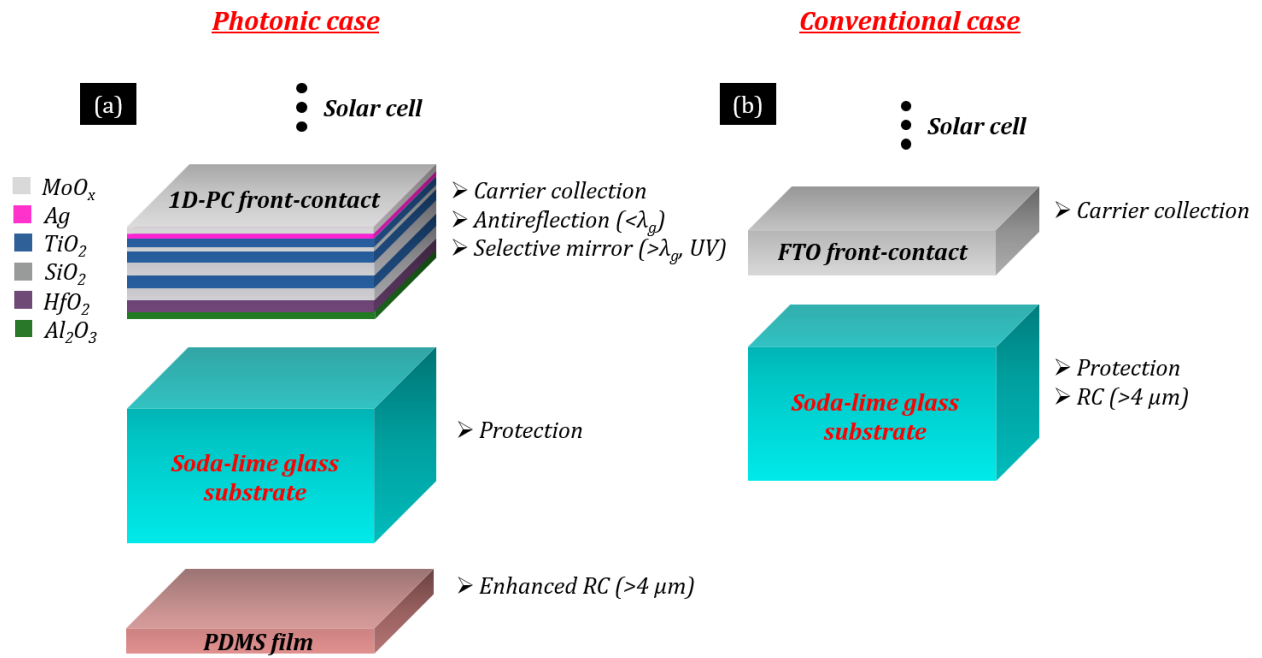

**Figure 2.** Multispectral-multifunctional photonic substrate for cooling solar cells. (a) Requirements for each spectral regime and the designed structure with a $Al_2O_3/HfO_2$/alternating $SiO_2/TiO_2/Ag/MoO_x$ photonic electrode for reflection of detrimental UV and sub-band gap radiation [near-infrared (NIR), short-wave-infrared (SWIR)] and enhanced transparency in visible, and a PDMS film for maximizing thermal radiation in mid-infrared (MIR). We note that in multilayer electrodes many other materials, i.e., instead of $TiO_2$ and $MoO_x$, may be utilized, such as all-$TiO_2$,[41] $SnO_2$, ZnO,[36] AZO,[35,39] or ultra-thin (<50 nm) ITO, GZO,[19] or FTO,[20] leading also to ultra-high transparency and low sheet resistance.[35,36,39,41] (b) Conventional TCO-covered glass substrate of solar cells with limited opto-electro-thermal response.

We perform a combined thermal-optical-electrical analysis to calculate the photonic cooling effect on PSCs operating temperature and efficiency as a function of the multilayer electrode's layers number, material, and thickness (see the Methods section) and an optimization procedure to optimize the multilayer electrode's structure and performance. The model considers all crucial macroscopic (i.e., sunlight absorption, emission, and nonradiative heat exchange) and microscopic processes (i.e., carrier generation, recombination and collection, as well as thermalization of hot generated carriers) of the cooler-cell system, in a broad wavelength range (e.g., 0.3–33 μm), i.e., considering both the thermal wavelengths (≥ 4 μm), including the atmospheric transparency window (8–13 μm), and the solar wavelengths (0.3–4 μm). First, we calculate the absorbed solar power in the encapsulated solar cell based on the simulated PSC absorptivity by employing the transfer matrix method and using it as the heat input in the electro-thermal simulation. We then set up a coupled electro-thermal simulator solving the steady-state energy balance for solar cells, with which we simulate the cell temperature ($T_c$) and PCE, assuming varying ambient temperature, humidity, and wind speed to mimic typical outdoor conditions (see the Methods section).

Due to light interference in the thin film stack, which strongly affects solar cell absorption properties, hence photocurrent and heat generation, we perform the multilayer electrode optimization on the total cooler-cell system. The optimization of the thin-film stack is implemented by combining a global evolutionary algorithm ("Genetic" algorithm) and a local optimization method ("Nelder Mead Simplex")[10] over an objective function, solar cells' PCE calculated according to the theoretical model (see the Methods section). The independent variables (optimized according to the objective function) are the thicknesses of the photonic electrode, photoactive layer, selective hole- (HTL), and electron-transporting (ETL) layers, while the glass substrate, FTO, MLG, and Au thicknesses are fixed to preserve low sheet resistance and high stability and mechanical protection. The examined PSC has a $SnO_2$ ETL instead of compact (c-) and mesoporous (m-) $TiO_2$, which require high annealing temperatures, low-cost multilayer Graphene back contact (MLG), and $FAPbI_3$ with enhanced thermal stability, transport properties, and optimal bandgap to reduce cost and enhance operational stability and PCE. We also demonstrate the high impact of our approach for the high-efficiency PSC case with Au back contact. The detailed structure of the device is as follows (see also Figure 1d): PDMS (15 μm)/TEC 8 (FTO – 500 nm)-covered glass substrate (FTO is replaced by a photonic engineered electrode in our case)/tin oxide ($SnO_2$ – 10–30 nm)/perovskite ($FAPbI_3$ – 350–800 nm)/Spiro-OMeTAD (150–200 nm)/multilayer Graphene (> ~10 μm) or Au (80 nm)/ethylene vinyl acetate (EVA – 0.47 mm)/polyvinyl fluoride (TEDLAR – 0.5 mm), where the numbers indicate the thickness of each layer, and the ranges indicate the boundary values of the independent variables corresponding to typical thicknesses in experiments. The detailed structure of the photonic electrode is as follows (see also Figure 2a): $Al_2O_3$ (20–200 nm)/$HfO_2$ (90–200 nm)/alternating $SiO_2$ (90–200 nm)/$TiO_2$ [90–200 nm, last $TiO_2$ (20–200 nm)]/Ag (8, 10, 12 nm)/$MoO_x$ (10–60 nm). Note that the Ag/$MoO_x$ thickness boundary values are lower to ensure low sheet resistance and high transparency.[35-38] To simulate the performance of our structures, we obtained the material parameters used for the active layer from ref 28, of the electrodes from refs 10,29–31, and of the other layers from refs 10,29,32.

Figure 3a–c shows the photonic cooling effect on PSCs' operating temperature ($T_c$ – red) and efficiency (PCE – blue) as a function of optimized photonic electrodes' $SiO_2$/$TiO_2$ sublayers number ($n$), assuming Ag thickness equals 8, 10, and 12 nm. Note that in all cases ($n$=2–9), the first sublayers correspond to $Al_2O_3$/$HfO_2$ for anti-reflection purposes (see Figure 2a). We also plot the cell operating temperature ($T_c$) and PCE of the conventional PSC (w/ FTO) and the PSC with a simple trilayer electrode (w/ $TiO_2$/Ag/$MoO_x$) for reference. We performed the optimization of the photonic electrodes assuming typical weather conditions, i.e., 1.7 m/s wind speed, 25 °C ambient temperature, and 40% relative humidity.[33] We also show later the impact of varying weather conditions. The red and blue inset numbers indicate the temperature decrease and the relative efficiency enhancement (%$_{rel.}$) compared to the conventional PSC case and correspond to the $TiO_2$/Ag/$MoO_x$ tri-layer electrode and the photonic multilayer electrode with the highest PCE improvement and the highest temperature reduction.

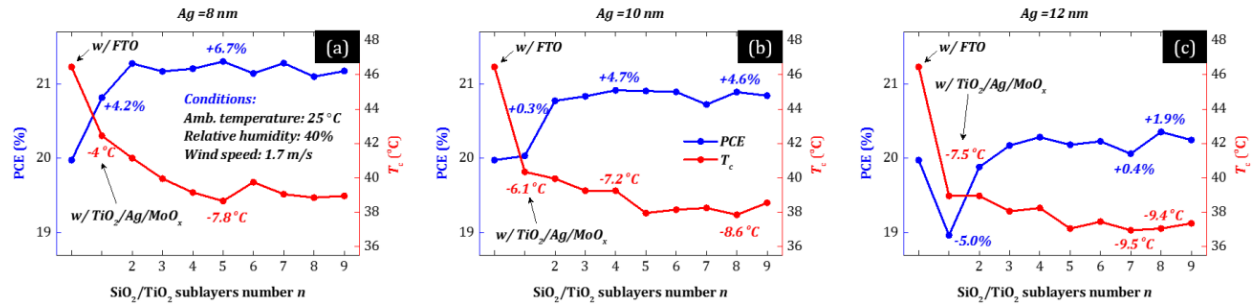

**Figure 3.** (a–c) Effect of the optimized photonic cooling substrate (see Figure 2a) on PSCs' w/ MLG operating temperature ($T_c$ – red) and power conversion efficiency (PCE – blue) as a function of optimized photonic electrodes' $SiO_2$/$TiO_2$ sublayers number ($n$), assuming Ag equals 8, 10, and 12 nm. $T_c$ and PCE of the conventional PSC case (w/ FTO) and the PSC case with a trilayer electrode (w/ $TiO_2$/Ag/$MoO_x$) are also plotted for reference. The optimization of the photonic electrodes was performed assuming typical weather conditions, i.e., 25 °C ambient temperature, 1.7 m/s wind speed, and 40% relative humidity.

First, in agreement with experimental results,[25,26] the PCE of conventional PSCs w/MLG is low (~20%), i.e., compared to the record PCEs of PSC configurations w/ Au (even beyond 25% for smaller cells),[22,23] mainly due to $V_{OC}$ loss due to the HTL-MLG interface (see also previous section and the left panel in Figure 1d).[26] Interestingly, as shown in Figure 3, the PCE of PSCs w/ MLG can significantly increase by applying the proposed photonic scheme. Specifically, as shown in Figure 3a, the PCE increases by 6.7%$_{rel.}$ (1.3% in absolute values), despite the compact 8 nm-thick Ag layer. Additionally, as shown in Figure 3a–c, the temperature of the PSC can significantly decrease by 7.8 to 9.5 °C (depending on Ag thickness). Such an improvement by over one percentage point of the absolute PCE is significant for PSCs since their PCE tends to saturate in recent years (since they are approaching the fundamental limit).[25] Especially, in the case of promising PSCs w/ MLG, i.e., of enhanced stability but lower PCE, finding ways to increase their PCE is of paramount importance. Moreover, a decrease in their operating temperature by over ~8 degrees during outdoor operation (with common $T_c$>50 °C) may significantly enhance their operational stability, leading to increased reliability/stability and higher system power output densities in the long term. Therefore, our work demonstrates a new technique for over one percentage point advance and over 8 degrees decrease in PSCs PCE and temperature, respectively, representing a significant advancement in PVs.

Interestingly, as shown in Figure 3a–b, a multilayer electrode consisting of only three layers ($TiO_2$/Ag/$MoO_x$) can provide substantial efficiency improvements (up to 4.2%$_{rel.}$) and temperature decrease (up to 6.1 °C) for a metal layer thickness of up to ~10 nm. Even a 12 nm-thick Ag on $MoO_x$ requires only four additional encapsulation layers to match the PCE of the conventional PSC case (see Figure 3c for $n$=2). Moreover, Figure 3 shows that the impact of adding more layers on PSCs' $T_c$ and PCE improvement increases with Ag thickness. Specifically, PCE increases by adding more layers compared to the trilayer electrode ($TiO_2$/Ag/$MoO_x$) by 2.5, 4.4, and 6.9%$_{rel.}$ for an Ag thickness of 8, 10, and 12 nm, respectively, indicating the importance of Ag encapsulation layers.

Figure 4 shows the physical origin of PSCs operating temperature decrease and PCE enhancement. Specifically, in Figure 4a–d, we plot the calculated solar reflectivity, transmissivity, and absorptivity/emissivity of the optimized photonic cooling substrate (PDMS/glass/photonic electrode – see Figure 2a), resulting in the highest PCE improvement when integrated into the PSC (see $n$=5 in Figure 3a). To effectively calculate the transmitted light into the perovskite layer in Figure 4, we also assume a $SnO_2$ thin film and a semi-infinite $FAPbI_3$ layer below the photonic electrode. Lines correspond to the ideal reflectivity, transmissivity, and absorptivity/emissivity spectra for optimum photonic cooling, plotted for reference, and the curves to the spectral response of the photonic cooling substrate with optimum design.

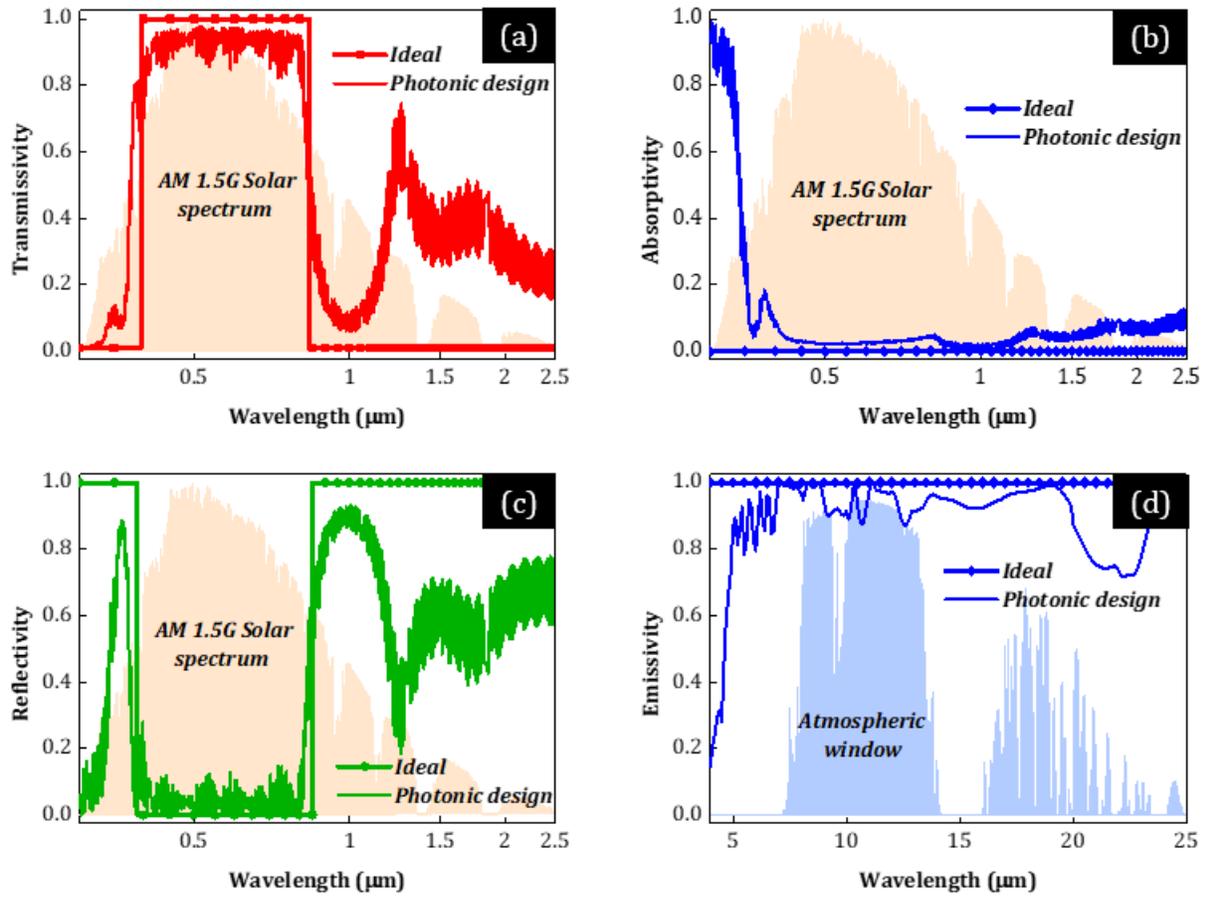

**Figure 4.** (a–d) Optimum (lines) solar reflectivity, transmissivity, absorptivity (0.3–2.5 μm), and thermal emissivity (>4 μm) spectra, respectively, for optimum photonic cooling of PSCs (i.e., with $\lambda_g \sim 0.832$ μm for FAPbI$_3$), compared to the ones calculated (curves) for the optimized photonic cooling substrate (PDMS/glass/photonic electrode – see Figure 2a) that results to the highest PSC PCE improvement when integrated into the PSC (see $n$=12 in Figure 3a), together with the AM1.5G solar irradiance spectra and the infrared transmission of the atmosphere.

The photonic cooling substrate provides broad-band spectral selectivity and high-performance filtering function from the sunlight to the infrared band. Specifically, it exhibits low absorption across the solar electromagnetic spectrum (0.3–4 μm), high reflection in the harmful UV (0.3–0.385 μm) and sub-band gap ($\lambda_g$-4 μm) spectral regimes, and high transparency in the visible. From our calculations, the average optical transmittance of the conventional FTO-covered glass substrate in the beneficial wavelength spectrum (0.385–$\lambda_g$ μm), which contributes to the photocurrent generation, is about 86.8 %, which agrees with commercial FTO (~84%).[42] Higher transmittance values often reported in the literature (>90%) mainly originate from the addition of common anti-reflection (AR) coatings at the air-glass interface, such as MgF$_2$ or LiF, or applying texturing.[22,23,43] Upon insertion of the multilayer electrode, the average transmittance increases to ~91.4% despite the compact 8 nm-thick Ag layer, an improvement of ~5.3%. This enhanced transmittance in the visible despite the reflective Ag layer is associated with reduced reflectance due to destructive interference of the reflected waves, where the reflectivity dip or transmission peak is controlled by the overcoat and undercoat layer thicknesses. In our study, the addition of a higher number of optimized encapsulation layers leads to broad-band transmission along the visible spectrum. Moreover, as seen in Figure 4c, the compact Ag layer provides broad-band reflection of the sub-band gap radiation, further optimized in the UV and near-IR regime by the multilayer. Specifically, due to the integrated multilayer electrode with Ag, essentially a 1D PC, the photonic cooling substrate reflects 70% of the sub-band gap radiation (calculated from the simulated reflectivity in $\lambda_g$-4 μm and the AM1.5G sun spectrum), compared to the conventional FTO-covered glass substrate, which only reflects 14.9%, a significant improvement of ~370%. Additionally, the photonic electrode-covered glass substrate parasitically absorbs 43.5 W/m$^2$ of solar power (calculated from the simulated absorptivity in 0.3–4 μm and the AM1.5G sun spectrum) compared to 134.0 W/m² of the conventional substrate due to the $R_{sh}$-$T$ trade-off mitigation (see Figure 4b), a significant improvement of ~68%. At mid-infrared wavelengths (>4 µm), the thermal emissivity is almost maximum (especially within the atmospheric window ~8–13 µm) due to the addition of the PDMS film placed on top of the glass front side, leading to black body-like radiative cooling. Specifically, the cooling power increases by 10 W/m² [from 100.0 to 110.0 W/m² – by solving Equations (2) and (3)], an improvement of 10%. We note that the photonic patterning of the PDMS layer may result in even higher cooling power and transmitted power in the visible, further enhancing solar cells' efficiency and decreasing their operating temperature.[12,44]

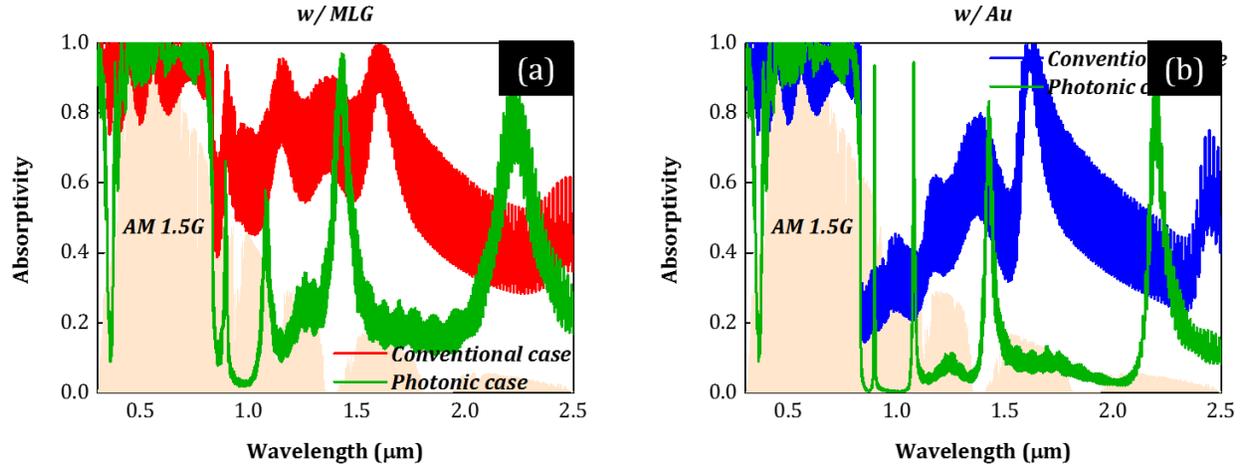

**Figure 5.** Solar absorptivity spectra of the PSCs (a) w/ MLG and (b) w/ Au in the photonic case (w/ PDMS/glass/photonic electrode - green) compared to the conventional PSC cases (w/ glass/FTO – red and blue, respectively).

With 91.4% visible light transmitted (~5.3% improvement), 70% infrared solar radiation reflected (~370% improvement), and 4.4% solar radiation absorbed (~68% improvement), integrating the photonic cooling substrate in the solar cell significantly modifies PSCs' spectral properties and improves their radiative response (see Figure 5). Specifically, Figure 5a and Figure 5b show the solar absorption of PSCs w/ MLG and w/ Au, respectively. The green curves correspond to the PSCs in the photonic case (i.e., w/ PDMS/glass/photonic electrode), and the red and blue curves to the conventional PSC cases (i.e., w/ glass/FTO). First, the reflection of the UV and sub-band gap radiation is significantly higher for the PSCs in the photonic case (green) than in the conventional cases (i.e., w/ FTO and Au – blue and w/ FTO and MLG – red). Additionally, the parasitic absorption of the PSCs in the photonic case is significantly lower than in the conventional cases across the solar spectrum. This improved optical response resulted in a significantly reduced PSC thermal load. Specifically, the thermal load, calculated by the total sun power (AM1.5G) minus the reflected and electrical power output of the PSCs at the steady state [by solving Equation (1)], was vastly reduced by 177.1 and 128.9 W/m² for w/ MLG (Figure 5a) and w/ Au cases (Figure 5b), respectively. The PSC optical response also significantly improved in the beneficial range of the solar spectrum (0.385–$\lambda_g$ µm – see Figure 5a and b). Specifically, the solar absorption in perovskite significantly increased (see Figure 5a) due to (i) the lower index contrast in the air-PDMS compared to the air-glass interface and (ii) enhanced anti-reflection at the glass-PSC interface induced by the optimized multilayer electrode.

As shown in Figure 6, PSCs' improved optical response translates to improved power conversion efficiencies due to improved external quantum efficiency (EQE – see Figure 6a and b), *J-V* characteristics (see Figure 6c and d) and power output from the solar cells (see Figure 6e and f). Specifically, Figures 6a and b, Figures 6c and d, and Figures 6e and f show the EQE, output current density, and power output, respectively, for an operating or cell temperature equal to the steady-state temperature calculated by solving Equation (1). Figures 6a, c, and e and Figures 6b, d, and f correspond to the PSCs w/ MLG and w/ Au, respectively. The green curves correspond to the PSCs in the photonic case (i.e., w/ PDMS/glass/photonic electrode), and the red and blue curves correspond to the conventional PSC cases (i.e., w/ glass/FTO).

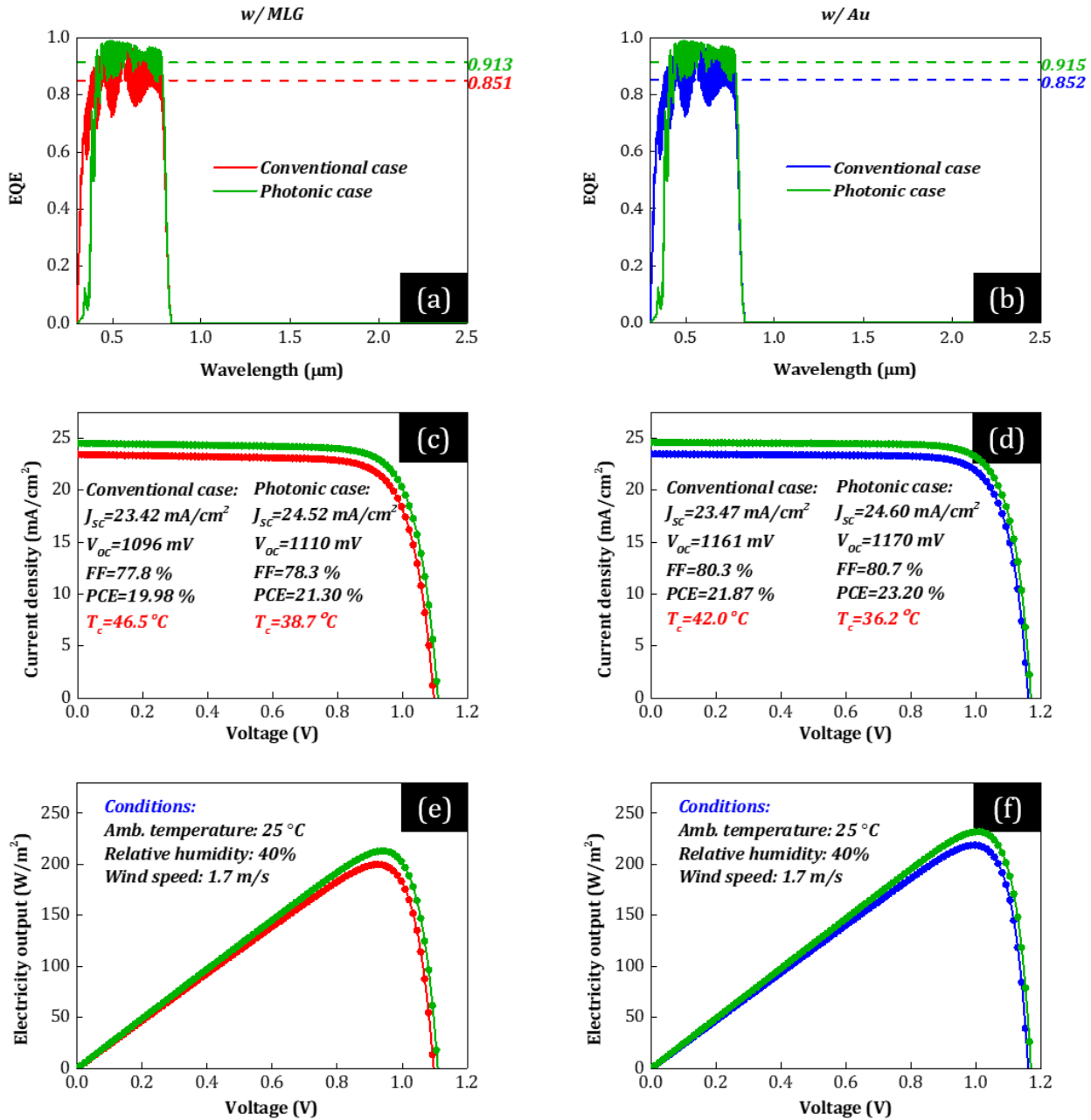

**Figure 6.** Effect of the optimized photonic cooling substrate (see Figure 2a) on (a, b) the external quantum efficiency (EQE) (c, d) the $J$–$V$ characteristics, and (e, f) power output of the PSCs w/ MLG (red – a, c, e), and w/ Au (blue – b, d, f), and the corresponding PV characteristics ($J_{SC}$, $V_{OC}$, FF, and PCE) and cell operating temperature ($T_c$), for common environmental conditions, i.e., 25 °C ambient temperature, 1.7 m/s wind speed, and relative humidity RH=40%. In (a–f), the green curves correspond to the PSCs in the photonic case (w/ PDMS/glass/photonic electrode – see also Figure 2a). The horizontal dashed lines in (a, b) correspond to the average EQE (at 0.4–0.8 μm) of each solar cell. All solar cell cases (conventional and photonic) correspond to the optimum devices optimized according to the optimization procedure discussed in the main text.

For both PSC cases, the output current density increases mainly due to the improved short-circuit current density ($J_{SC}$). Specifically, the calculated $J_{SC}$ obtained by integrating the EQE over the AM1.5G spectrum [see Equation (8) and Figure 6a and b] improves from 23.42 mA/cm² to 24.52 mA/cm² and from 23.47 mA/cm² to 24.60 mA/cm² (an increase of 4.7% and 4.8% in relative terms) for w/ MLG (Figure 6c) and w/ Au cases (Figure 6d), respectively. The $J_{SC}$ increase is due to the enhanced solar absorption in the active layer due to improved impedance matching and anti-reflection induced by the PDMS film and the photonic electrode in the beneficial range of the solar spectrum (0.385–$\lambda_g$ μm – see also Figures 4 and 5). (We note that the EQE and hence $J_{SC}$ in the conventional and photonic cases can be even higher with the addition of common AR coatings on substrate's front side, such as $MgF_2$ or LiF,[22,23,43] or applying photonic patterning.[12,44]) Moreover, despite the increased solar absorption in 0.385–$\lambda_g$ μm (and hence the associated increased parasitic and thermalization losses that result in higher heat dissipation in the structure), at the steady state [by solving Equation (1)], the enhanced cooling effect and the reduced parasitic heat generation of the PSCs in the photonic case result in lower operating temperatures than the conventional PSC cases (see insets in Figure 6c and d), which further affects the $J$–$V$ characteristics. Specifically, with a decrease in the operating temperature of the PSC, the carrier concentration decreases exponentially [see Equation (11)], leading to a lower dark current density [see Equation (10)]. Open-circuit voltage, $V_{OC}$, [i.e., solving Equation (7) for $J$=0] increases exponentially with a decrease in the dark current at lower temperatures, leading to a temperature impact on PSCs' energy yield and a negative PSC power-temperature coefficient ($\beta$ – see Table 1). Therefore, the output voltage also increases due to the temperature reduction and the $V_{OC}$ rise, which is more evident in the case of the PSC w/ MLG due to the higher temperature decrease (see insets and green versus red, blue lines in Figure 6c and d, respectively, for $J$=0). Eventually, due to the $J_{SC}$, $V_{OC}$, and FF increase, the power conversion efficiency [PCE($V_{mp}$,$T_c$)=$J_{SC}V_{OC}(T_c)FF(T_c)/\int I(\lambda)d\lambda$=$J(V,T_c)V(T_c)|_{mp}/\int I(\lambda)d\lambda$] increases by ∼6.6% and ∼6.1%, in relative terms, for the PSCs w/ MLG and w/ Au, respectively, assuming typical weather conditions, i.e., 1.7 m/s wind speed, 25 °C ambient temperature, and 40% humidity.

We examine the impact of varying environmental conditions on photonic cooling in PSCs in the following figures (Figures 7, 8, and 9). Specifically, the main challenge for the practical deployment and commercialization of PSCs is their operational stability, which is significantly affected by the cell temperature.[4,5] Environmental conditions (i.e., wind speed, ambient temperature, and humidity) play a critical role in the operating temperature of solar cells. Indicatively, under natural conditions of outdoor operation, the device can easily reach temperatures higher than 50 °C, sometimes (depending on the weather conditions) even reaching values close to the lamination-fabrication temperatures, which are carefully selected to be the lowest possible (<85 °C).[1] This heating of the PSC has adverse consequences on its PCE and reliability. Therefore, a photonic approach should provide cell temperature reductions and PCE enhancements irrespective of the environmental conditions. Especially in extreme environmental conditions, i.e., high ambient temperature, humidity, and low wind speed, the temperature decrease should be as high as possible to avoid permanent damage of the PSC. To examine the impact of varying environmental conditions on photonic cooling in PSCs, we plot in Figures 7, 8, and 9 the cell temperature ($T_c$ – a, b) and power conversion efficiency (PCE – c, d) of PSCs w/ MLG (a, c) and w/ Au (b, d), as a function of wind speed, ambient temperature, and relative humidity (RH), respectively. The green curves correspond to the PSCs in the photonic case (i.e., w/ PDMS/glass/photonic electrode), and the red and blue curves to the conventional PSC cases (i.e., w/ glass/FTO).

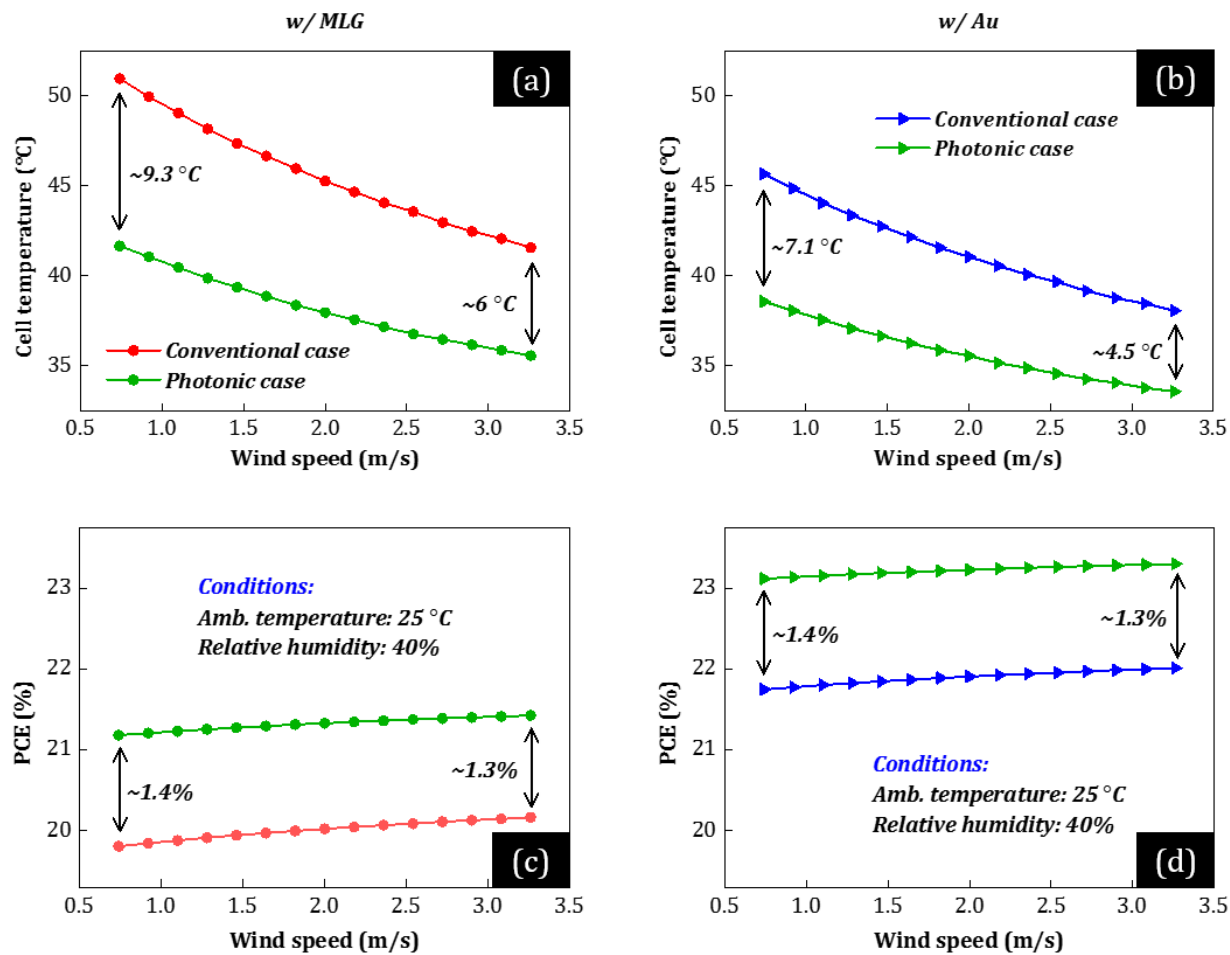

**Figure 7.** (a, b) Cell operating temperature, $T_c$, and (c, d) power conversion efficiency, PCE, of PSCs w/ MLG (a, c) and w/ Au (b, d), as a function of wind speed, assuming 25 °C ambient temperature and relative humidity RH=40%. The green curves in (a–d) correspond to the PSCs in the photonic case (i.e., w/ PDMS/glass/photonic electrode). The red and blue curves in (a–d) correspond to the conventional PSC cases (i.e., w/ glass/FTO).

In Figure 7, we assume varying wind speeds in the range of 0.75–3.25 m/s, constant ambient temperature $T_a$=25 °C, and relative humidity RH=40%. First, our theoretically calculated operating temperatures of conventional PSCs (see red and blue curves) may reach values up to ~50 °C, in agreement with experimental studies,[1] comparable to bulkier silicon-based counterparts (∼250 μm thickness).[33] Moreover, the conventional PSC w/ MLG (red) operates at 3.5 to 5.3 °C (depending on wind speed) higher cell temperatures than w/ Au (blue). The reason is the increased solar heating power, originating from the increased parasitic absorption in the MLG back contact, mainly in the sub-band gap regime (see also Figure 1b and Figure 5a versus b). Additionally, the conventional PSC w/ MLG (red) exhibits lower absolute efficiencies by up to 3% than w/ Au (blue), mainly due to open-circuit voltage ($V_{OC}$) loss due to the HTL-MLG interface (see also right panel in Figure 1a). These results demonstrate the importance of lowering the operating temperature and enhancing the PCE of PSCs, especially of promising PSCs with MLG-based back contacts. Notably, the operating temperature of the PSCs in the photonic case (green) is up to 9.3 and 7.1 °C lower than that in the conventional cases w/ MLG (red) and w/ Au (blue), respectively, a remarkable improvement compared to approaches that appear in the literature.[8,13,14,45,46] Such a temperature decrease may significantly enhance the operational stability of PSCs, leading to increased reliability/stability and higher system power output densities in the long term. Especially in extreme environmental conditions, i.e., low wind speed, where the cell temperature of sensitive material-based solar cells such as PSCs should be as low as possible to avoid permanent damage, the cell temperature decrease is the highest and over 9 °C. Moreover, as shown in Figure 7b, the absolute PCE of the PSCs in the photonic case is significantly higher than that in the conventional cases, i.e., up to 1.4%. Most importantly, this PCE improvement is insensitive to varying wind speed mainly due to improved solar absorption of visible light in the active layer (see also Figures 4, 5, and 6) provided by the integrated photonic scheme.

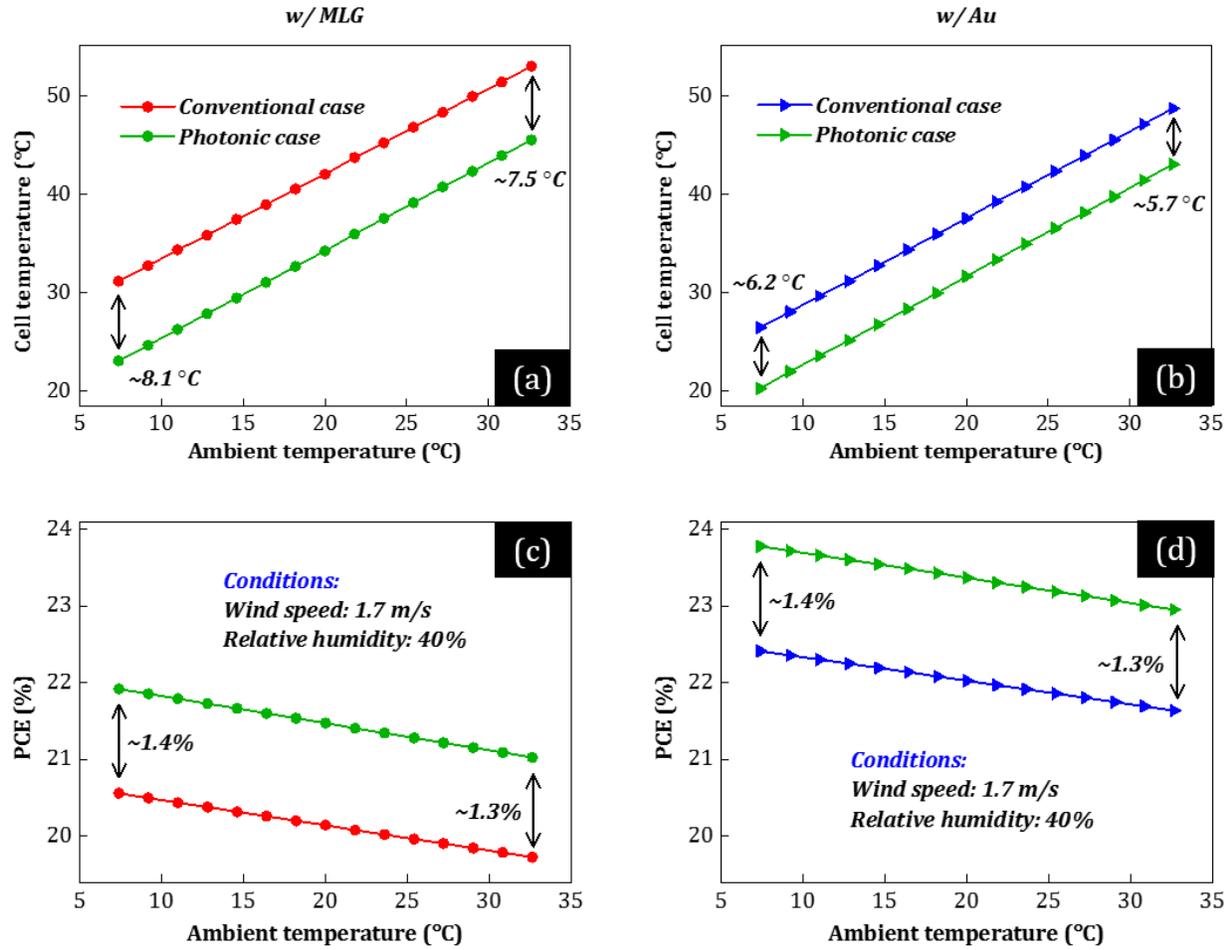

**Figure 8.** (a, b) Cell operating temperature, $T_c$, and (c, d) power conversion efficiency, PCE, of PSCs w/ MLG (a, c) and w/ Au (b, d), as a function of ambient temperature, assuming 1.7 m/s wind speed and relative humidity RH=40%. The green curves in (a–d) correspond to the PSCs in the photonic case (i.e., w/ PDMS/glass/photonic electrode). The red and blue curves in (a–d) correspond to the conventional PSC cases (i.e., w/ glass/FTO).

Figures 8 and 9 also demonstrate the insensitivity of the PCE enhancement and the cell temperature reduction to varying environmental conditions, where we plot PSCs' cell temperature and PCE as a function of ambient temperature and relative humidity, respectively. We assume varying ambient temperature in the range of 7.5–32.5 °C, constant wind speed of 1.7 m/s, and relative humidity RH=40% in Figure 8, and varying humidity in the range of 50–100%,[47] constant ambient temperature of 30 °C, and wind speed of 1.7 m/s in Figure 9. In all cases (see Figure 8 and 9), i.e., no matter ambient temperature and humidity, the operating temperature of the PSCs in the photonic case (green) is 7.3 to 8.5 °C and 5.4 to 6.2 °C lower than that in the conventional cases w/ MLG (red) and w/ Au (blue), respectively. This insensitivity of cell temperature to varying environmental conditions is due to enhanced selective-spectral and radiative cooling.[7] Additionally, in all cases, the PCE of the PSCs in the photonic case (green) is 1.3 to 1.4% higher than that in the conventional cases. This insensitivity of PCE to the different cell materials and varying environmental conditions is mainly due to improved solar absorption of visible light in the active layer provided by the integrated photonic scheme (see also Figures 4, 5, and 6) for all-weather, high performance.

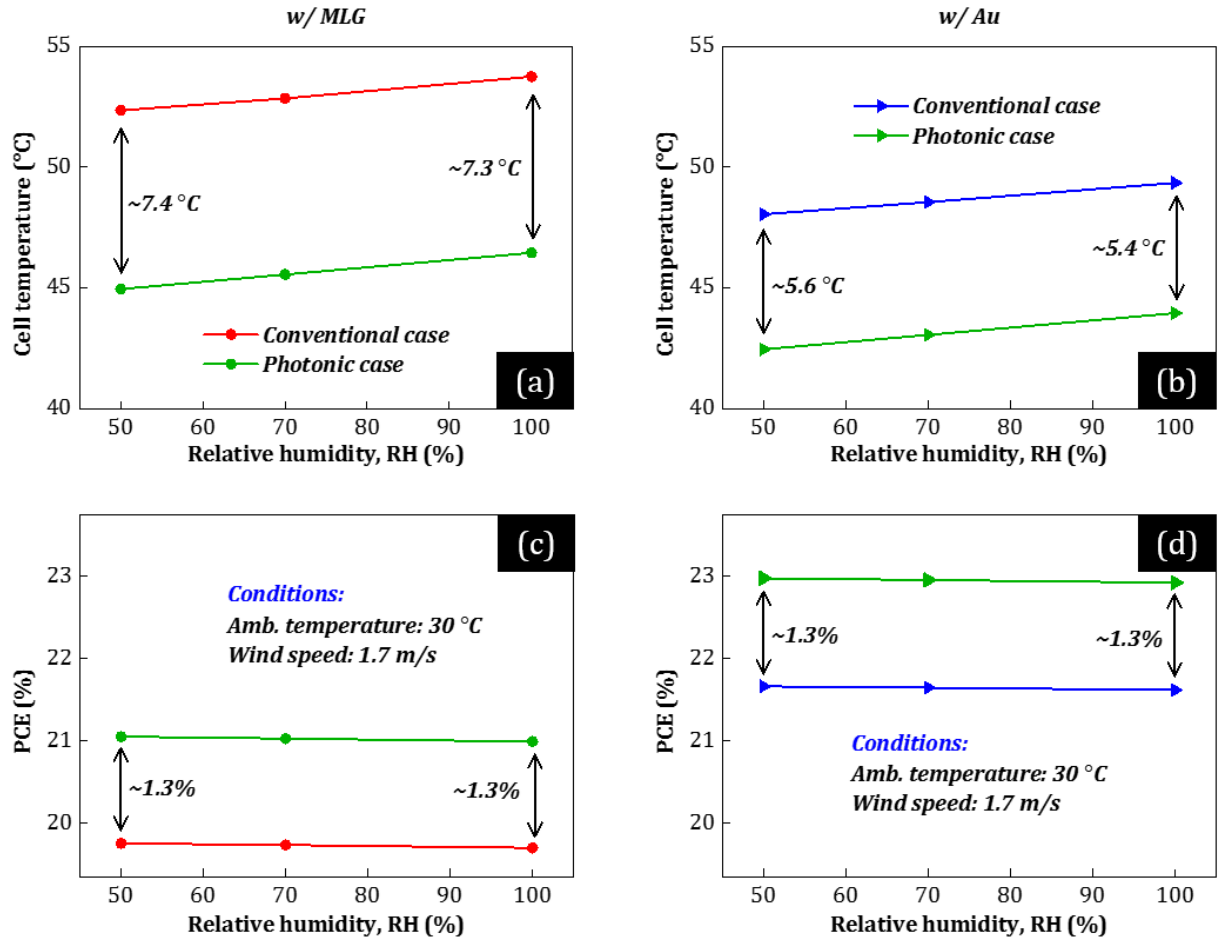

**Figure 9.** (a, b) Cell operating temperature, $T_c$, and (c, d) power conversion efficiency, PCE, of PSCs w/ MLG (a, c) and w/ Au (b, d), as a function of humidity, RH, assuming 25 °C ambient temperature and 1.7 m/s wind speed. The green curves in (a–d) correspond to the PSCs in the photonic case (i.e., w/ PDMS/glass/photonic electrode). The red and blue curves in (a–d) correspond to the conventional PSC cases (i.e., w/ glass/FTO).

■ **CONCLUSIONS**

We have introduced a multispectral-multifunctional photonic concept for cooling solar cells and advancing current PV performance with high-efficiency, low-cost, and sensitive next-generation solar cell materials, such as perovskites. In particular, we propose to cover the glass substrate of solar cells with a cost-effective multifunctional photonic structure and a PDMS film. The polymer film maximizes solar cells' thermal radiation, enhancing radiative cooling. The photonic structure, which is a 1D (regarding periodicity) multilayer stack, reflects useless solar radiation, reducing the parasitic heat generation. Additionally, the photonic structure simultaneously serves as an ultra-thin transparent front contact and an anti-reflection coating, further reducing the parasitic heat generation and enhancing efficiency. To determine the potential of our photonic approach and fully exploit the cooling potential of PSCs, we first experimentally characterized the thermal radiation and solar absorption properties of the most promising PSC configurations (regarding efficiency, stability, and cost). Then a coupled opto-electro-thermal simulation was conducted to investigate the effect of our approach on PSCs operating temperature and PCE based on promising PSC configurations, as a function of varying environmental conditions. The results show that our approach can efficiently reduce the cell operating temperature and significantly enhance the PCE. Although we have focused on perovskite-based photovoltaics in terrestrial environments, our approach can, in principle, be applied (with even higher impact) to PSCs in space applications, concentrated PSCs, other solar cells with optimal band gap, hybrid thermoelectric-photovoltaic systems, heat-insulating schemes, or other energy-saving applications, such as smart windows.

Therefore, our study paves the way to novel photonic multifunctional and multispectral electrodes that can affect the opto-electro-thermal response of a device.

■ **METHODS**

**Perovskite Solar Cells Fabrication.**

Fluorine-doped tin oxide (FTO) glass substrates (TCO glass, TEC8) were etched using Zn powder and diluted hydrochloric acid (HCl), cleaned by ultrasonication in Hellmanex (2%, deionized water), deionized water, acetone, and ethanol. After drying the substrates with a nitrogen gun, they were UV-$O_3$ treated for 15 min. Afterwards, an approximately 20 nm thick blocking layer ($TiO_2$) was deposited on the FTO by spray pyrolysis at 450 °C using a commercial titanium diisopropoxide bis(acetylacetonate) solution (75% in 2-propanol, Sigma-Aldrich) diluted in anhydrous ethanol (1:9 volume ratio) as a precursor and oxygen as a carrier gas. A mesoporous $TiO_2$ layer was deposited by spin-coating a diluted paste (Dyesol 30NRD) in ethanol (1:6 weight ratio) at 4000 rpm for 15 s and sintering at 450 °C for 30 min in a dry-air atmosphere. The perovskite films were deposited from the precursor solution, which was prepared in an argon atmosphere by dissolving FAI, MABr, $PbI_2$ and $PbBr_2$ in anhydrous dimethylformamide/dimethyl sulfoxide (4:1 volume ratio) to achieve the desired compositions $(FAPbI_3)_{0.98}(MAPbBr_3)_{0.02}$ using a 3% $PbI_2$ excess and 44 mg of MACl. The perovskite precursor was deposited in a dry-air atmosphere on FTO/c-$TiO_2$/m-$TiO_2$ substrate, using a single-step deposition method (6000 rpm for 50 seconds). To control the film crystallization, 10 seconds before the end of the spin-coating program, the perovskite precursor was quenched with chlorobenzene as the antisolvent. To form and crystallize the perovskite, the spin-coated perovskite precursors were annealed at 150 °C for 30 minutes inside a dry-air atmosphere. Subsequently, the perovskite films were then passivated by spin-coating (6000 rpm for 50 s) a 3 mg mL$^{-1}$ dispersion of octylammonium iodide (OAI) in isopropanol. The HTM (Spiro-OMeTAD doped with bis(trifluoromethylsulfonyl)imide lithium salt (17.8 μL of a solution of 520 mg of LiTFSI in 1 mL of acetonitrile) and 28.8 μL of 4-*tert*-butylpyridine)) was deposited by spin-coating at 4000 rpm for 30 s. Finally, an approximately 80 nm gold (Au) layer or 10 μm multi-layer Graphene layer (homogenized Timrex KS25 powders), were deposited by thermal evaporation and doctor-blade coating, respectively.

**Perovskite Solar Cells Characterization.**

Fourier-transform infrared spectroscopy (FT-IR) measurements were carried out under vacuum, with a Bruker Vertex 70v FT-IR vacuum spectrometer (Bruker Optik GmbH, Rosenheim, Germany); The transmission of the samples was evaluated using a PIKE universal sample holder (PIKE Technologies, Inc. – Madison, USA), while reflection was measured using a Bruker Optics A513 reflection accessory (Bruker Optik GmbH, Rosenheim, Germany), at an angle of incidence of 7-degrees. To cover a spectral range of 0.45–25 μm, two different sets of optics were used: (a) for 0.45–1.25 μm, a Quartz beamsplitter and a room temperature Silicon diode detector, while (b) for 1.3–25 μm), a broad band KBr beamsplitter and a room temperature broad band triglycine sulfate (DTGS) detector were used. In any case, interferograms were collected at 4 cm$^{-1}$ resolution (8 scans), apodized with a Blackman-Harris function, and Fourier transformed with two levels of zero filling to yield spectra encoded at 2 cm$^{-1}$ intervals. Before scanning the samples, an empty holder and an aluminum mirror (>90% average reflectivity) background measurement was recorded in vacuum for transmission and reflection measurements, respectively, and each sample spectrum was obtained by automatic subtraction of it.

**Opto-Electro-Thermal Calculation of Perovskite Solar Cells.**

We perform a combined thermal-optical-electrical analysis to calculate the photonic cooling effect in PSCs. First, we calculate the absorbed solar power in the encapsulated solar cells based on the simulated PSCs absorptivity, by employing the transfer matrix method, and use it as the heat input in the electro-thermal simulation. We then set up a coupled electro-thermal simulator solving the steady-state energy balance for solar cells, with which we simulate the cell operating temperature ($T_c$) and the power conversion efficiency (PCE), assuming varying ambient temperature, humidity, and wind speed to mimic typical outdoor conditions:[10,11]

$$P_r(T_c) + P_c(T_c, T_a) + P_g(T_c, T_a) = P_h(V_{mp}, T_c) + P_a(T_a), (1)$$

In Equation (1), $P_h(T_c)$ is the heat flux from solar radiation and $P_a(T_a)$ is the radiative heat flux from the atmosphere, absorbed by the device at ambient temperature, $T_a$. $P_r(T_c)$ is the total heat flux radiated by the solar cell at $T_c$, $P_c(T_c, T_a)$ accounts for the outgoing nonradiative heat transfer, and $P_g(T_c, T_a)$ is the radiative heat flux by the rear surface of the solar cell. These power terms are given by[7,10]

$$P_r(T_c) = \int_0^\infty \int_0^{2\pi} \int_0^{\pi/2} I_{BB}(\lambda, T_c)\varepsilon(\lambda, \theta, \varphi)\cos\theta\sin\theta d\theta d\varphi d\lambda, \quad (2)$$

$$P_a(T_a) = \int_0^\infty \int_0^{2\pi} \int_0^{\pi/2} I_{BB}(\lambda, T_a)\varepsilon(\lambda, \theta, \varphi)\varepsilon_a(\lambda, \theta)\cos\theta\sin\theta d\theta d\varphi d\lambda, \quad (3)$$

$$P_g = \sigma\varepsilon_r A(T_c^4 - T_a^4), \quad (4)$$

$$P_c(T_c, T_a) = h_c(T_c - T_a), \quad (5)$$

$$P_h(V_{mp}, T_c) = \int_0^\infty I(\lambda)\varepsilon(\lambda)d\lambda - PCE(V_{mp}, T_c)\int_0^\infty I(\lambda)d\lambda, \quad (6)$$

where $\lambda$ is the free-space wavelength, $\sigma$ is the Stefan-Boltzmann constant, $A \sim 1$ is the view factor, $I_{BB}(\lambda, T_c)$ is the spectral intensity of a blackbody at temperature $T_c$ given by Planck's law, $I(\lambda)$ is the solar illumination represented by the measured sun's radiation, the AM1.5G spectrum, and $h_{c,top}$ and $h_{c,bottom}$ are the wind-speed-dependent nonradiative heat transfer coefficients (higher $h_c$ values correspond to stronger winds) at the top and rear surfaces of the solar cell, respectively. For $h_{c,top}$ and $h_{c,bottom}$, we use two relations, frequently used in previous studies for similarly encapsulated solar cell systems, expressed as $h_{c,top} = 5.8 + 3.7v_w$ and $h_{c,bottom} = 2.8 + 3.0v_w$, where $v_w$ is the velocity of wind at the module surface (in m/s) given by the relationship suggested in the literature $v_w = 0.68v_f - 0.5$, where $v_f$ is the wind speed measured by the closest weather station.[33] $\varepsilon(\lambda, \theta, \varphi)$ is the solar cell spectral directional emissivity (equal to absorptivity, according to Kirchhoff's law), $\varepsilon_a(\lambda, \theta) = 1 - t(\lambda)^{1/\cos\theta}$ is the angle-dependent emissivity of the atmosphere, with $t(\lambda)$ the atmospheric transmittance in the zenith direction, and $\varepsilon_r \sim 0.85$ is the solar cell rear surface hemispherical emissivity.[10] Due to energy conservation, $P_h$ equals the difference between absorbed solar energy flux and generated electrical power in the solar cell, where $PCE(V_{mp},T_c)=J(V,T_c)V(T_c)|_{mp}/\int I(\lambda)d\lambda$ is the temperature-dependent cell's solar-to-electrical power conversion efficiency (PCE) assuming that it operates at its maximum power point (mp),[10] where $J$ and $V$ are the output current density and voltage, respectively. In Equation (6), we assume that the structure is facing the sun at a fixed angle. Thus, the term $P_h$ does not have an angular integral, and solar cell's absorptivity/emissivity is represented by its value at normal incidence.

In the present study, we assume dominating recombination by the space charge region since most of the perovskite layer is depleted.[48] Assuming that Shockley-Read-Hall recombination is the dominant nonradiative recombination mechanism and in the presence of shunt resistance (accounting for manufacturing defects and impurities near the junction), we calculate the current–voltage characteristics by the following diode equation:[7,10,49]

$$J(V, T_c) = J_{SC} - J_{r,0}(T_c)\left(e^{(qV/k_BT_c)} - 1\right) - J_{nr,0}(V, T_c)\left(e^{(qV/2k_BT_c)} - 1\right) - \frac{V}{R_{sh}}, \quad (7)$$

where $q$ is the elementary charge of an electron, $k_B$ is Boltzmann's constant, and $R_{sh}$ is the solar cell shunt resistance. The term

$$J_{SC} = q\int_0^{\lambda_{BG}} I(\lambda)EQE(\lambda)d\lambda, \quad (8)$$

is the current density flowing at short-circuit conditions under the sun illumination, where $EQE(\lambda)$ is the external quantum efficiency of the solar cell. $EQE(\lambda)$ is defined as the multiplication of the internal quantum efficiency [i.e., number of charge carriers collected versus the number of absorbed photons – $IQE(\lambda)$] and the absorption efficiency of the active layer, $\varepsilon_{al}(\lambda)$, i.e., $EQE(\lambda)=IQE(\lambda)\varepsilon_{al}(\lambda)$. $IQE(\lambda)$ is extracted from ref 50 and is close to unity, mainly due to the low thickness of the active layer. The second and the third terms correspond to the radiative and nonradiative recombination current densities, respectively, with the corresponding dark-saturation current densities $J_{r,0}$ and $J_{nr,0}$, given by Equations (9) and (10), and ideality factors of 1 and 2, respectively:

$$J_{r,0}(T_c) = q\int_0^{\lambda_{BG}} I_{BB}(\lambda)\varepsilon_{al}(\lambda)\,d\lambda, \quad (9)$$

$$J_{nr,0}(V, T_c) \propto \frac{n_i(T_c)}{\tau}\sqrt{(V_{bi}(T_c) - V)}, \quad (10)$$

where $\tau = 1$ μs is the lifetime of electrons and holes (assuming equal lifetimes for electrons and holes) extracted from ref 51, $V_{bi}(T_c)$ is the temperature-dependent built-in bias, and $n_i(T_c)$ is the temperature-dependent intrinsic charge carrier density. Since the built-in bias is typically slightly higher than the open-circuit voltage, we set it a bit higher than the open-circuit voltage of the Shockley-Queisser limit for perovskite material's bandgap.[49] The density of states in the conduction, $N_C$, and the valence band, $N_V$, assuming $N_C=N_V$, are extracted by DFT calculations.[52] We calculate then the temperature-dependent intrinsic charge carrier density by

$$n_i(T_c) = 2\left(\frac{2\pi m k_B T_c}{h^2}\right)^{3/2} e^{-E_g(T_c)/2k_B T_c}, (11)$$

where $h$ is Planck's constant, $m$ is the effective mass of the electrons and holes (assuming equal electrons', holes' effective mass), and $E_g(T_c)$ is the temperature-dependent bandgap, which is assumed to increase by 0.35 meV per 1 K.[53] We fit our model to the current–voltage characteristics of the examined promising PSCs w/ MLG and w/ Au back contacts.[23,26] The PV characteristics at 25 °C and 1000 W/m² of solar radiation, i.e., short-circuit current density, $J_{SC}$, open-circuit voltage, $V_{OC}$, (i.e., for $J=0$), fill factor, $FF=J(V)V|_{mp}/J_{SC}V_{OC}$, and output power-temperature coefficient ($\beta$) are summarized in Table 1.

| PSC | $V_{OC}$ (V) | $J_{SC}$ (mA/cm²) | $FF$ (%) | PCE (%) | $\beta$ (%/°C) |
|---|---|---|---|---|---|
| w/ MLG | 1.12 | 22.66 | 78.9 | 20.06 | -0.25 |
| w/ Au | 1.18 | 25.50 | 81.2 | 24.53 | -0.25 |

To evaluate the validity of the modeled solar cell's temperature dependence, we compare our calculated power-temperature coefficients [i.e., the slopes of the $PCE$ (%)–$T_c$ curves] with those in literature [the slopes of the $PCE$ (%)–$T_c$ curves are normalized at % compared to the solar cells operating at Standard Test Conditions (STC) (i.e., 1000 W/m² of solar radiation, $T_c=298.15$ K)]. The calculated power-temperature coefficients ($\beta$) of the next-generation perovskite-based solar cells are equal to -0.25 %/°C, in agreement to literature reports calculated from experimental data for solar cells' typical operating temperatures range.[1] Interestingly, they are also much lower than conventional silicon-based solar panels (around -0.4 %/°C), implying superior performance at high-temperature conditions.

**Notes**

The authors declare no competing financial interest.

## ■ ACKNOWLEDGMENTS

GP acknowledges support by the Hellenic Foundation for Research and Innovation (HFRI) under the "2nd Call for H.F.R.I. Research Projects to Support Faculty Members & Researchers", Project ID 2936 (META-ENERGY). G.K. acknowledges financial support from the European Commission's Marie Skłodowska-Curie Actions, H2020-MSCA-IF-2020, project ID 101024237.

## ■ REFERENCES